\documentclass[twocolumn,longbibliography]{revtex4-1}
\usepackage{graphicx}
\usepackage{physics}
\usepackage{mathrsfs}
\usepackage{hyperref}
\usepackage{mathtools}
\usepackage{tabularx}

\usepackage{amsmath,amsfonts,amssymb,amsthm}

\hypersetup{
 linktocpage=true,
 pdfborderstyle={/S/S/W 1},
 hyperindex=true,
 bookmarks=true,
 bookmarksopen=true,
 bookmarksnumbered=true,
}

\begin{document}

\title{Gaussian multipartite quantum discord from classical mutual information}
\author{Mark Bradshaw, Ping Koy Lam, Syed M. Assad}
\affiliation{Centre for Quantum Computation and Communication Technology, Department of Quantum Science,\\ Research School of Physics and Engineering, Australian National University, Canberra ACT 2601, Australia.}

\begin{abstract}
  Quantum discord is a measure of non-classical correlations, which are excess correlations inherent in quantum states that cannot be accessed by classical measurements. For multipartite states, the classically accessible correlations can be defined by the mutual information of the multipartite measurement outcomes. In general the quantum discord of an arbitrary quantum state involves an optimisation of over the classical measurements which is hard to compute. In this paper, we examine the quantum discord in the experimentally relevant case when the quantum states are Gaussian and the measurements are restricted to Gaussian measurements. We perform the optimisation over the measurements to find the Gaussian discord of the bipartite EPR state and tripartite GHZ state in the presence of different types of noise: uncorrelated noise, multiplicative noise and correlated noise. We find that by adding uncorrelated noise and multiplicative noise, the quantum discord always decreases. However, correlated noise can either increase or decrease the quantum discord. We also find that for low noise, the optimal classical measurements are single quadrature measurements. As the noise increases, a dual quadrature measurement becomes optimal.
\end{abstract}

\maketitle

\section{Introduction}

A pair of quantum systems can be
\emph{entangled}~\cite{horodecki2009quantum}. Entangled quantum states
posses a form of correlation not possible with classical systems. If two
quantum states are not entangled, they are said to be \emph{separable}. Separable quantum states can be created
through local operations and classical communication. However,
separable quantum states can still possess correlations that are not
accessible through local measurements~\cite{bennett1999quantum}. Quantum discord (QD) was proposed by Ollivier and
Zurek~\cite{ollivier2001quantum} and Henderson and
Vedral~\cite{henderson2001classical} as a means of quantifying the
quantum correlations present in bipartite states that are not
necessarily entangled. To quantify the locally accessible (classical)
correlations, this quantification involves a measurement on one of the
subsystem. This measurement is chosen to maximize the classical
correlations. In general, this quantum discord will be different
depending on which subsystem is measured. As such, we will refer to
this as the {\it asymmetric} QD.

A desirable property of correlations might be for them to be symmetric
and one way to impose this property is to require that both parties
measure their subsystems. Such symmetric versions of the
quantum discord have been proposed; the {\it symmetric} QD is defined
by requiring a projective measurements of both
subsystems~\cite{maziero2010symmetry}. Alternatively, another version
of the QD can be defined involving arbitrary measurements on each
subsystem~\cite{piani2008no, wu2009correlations,
  terhal2002entanglement}, we call this the {\it extended} symmetric
QD.

QD can also be extended to more than two parties. The
multipartite symmetric QD quantifies the correlations present when
there are three or more parties, and when each party performs
projective measurements on their subsystem~\cite{rulli2011global}. It
can also be defined for the situation in which each party performs
arbitrary measurements~\cite{piani2008no}, which we call the
multipartite extended symmetric QD.

Calculating the asymmetric QD is an NP-hard
problem~\cite{1367-2630-16-3-033027}. The symmetric QD, and extended
symmetric QD, and their multipartite extensions, are likely just as
difficult. For continuous variable states, one can consider Gaussian
versions of QD. If the state is Gaussian, restricting the measurement
to Gaussian measurements give rise to the Gaussian
QD~\cite{giorda2010gaussian,adesso2010quantum}. This restriction
significantly reduces the number of variables involved in the
optimisation for finding the optimal measurement. The Gaussian discord is
asymmetric as it involves a measurement on only one of the subsystems. In
this paper, we define and investigate the symmetric and multipartite
versions of the Gaussian QD.

There are many other ways of defining quantum discord-like
measures. The quantum discord can be defined as the distance to the
closest classical state in terms of relative
entropy~\cite{modi2010unified}, or trace distance which gives the
geometric quantum discord~\cite{dakic2010necessary}. The quantum work
deficit~\cite{oppenheim2002thermodynamical} describes the difference
in work that can be extracted from a heat bath if one party is in
possession of bath subsystems compared to when they are
not. Measurement-induced nonlocality~\cite{luo2011measurement}
quantifies the distance between the pre and post measurement state,
when a local projective measurement is performed on one subsystem
without disturbing the subsystem. The interferometric
power~\cite{qantum2014girolami} quantifies how helpful a quantum state
is for estimating a parameter of a Hamiltonian that acts on one of the
subsystems. See~\cite{quantum2017bera} for a review of quantum discord
measures.

This paper is organised as follows: In section \ref{sec_background}, we describe the asymmetric QD, the
symmetric QD, extended symmetric QD, and multipartite extended
symmetric QD. In section \ref{sec_gaussian_multi}, we introduce the
Gaussian multipartite QD, describe its properties, and calculate it
for a two-mode EPR state and a three-mode tripartite GHZ state
subjected to different types of noise. Finally, we summarize our
results in section \ref{sec_conclusion}.

\section{Background}
\label{sec_background}

\subsection{Asymmetric quantum discord}

The total correlations present in a bipartite quantum state $\rho_{AB}$ is given by the quantum mutual information (MI),
\begin{equation}
I_Q(A;B) = S(A) + S(B) - S(AB),
\end{equation}
where $S$ is the von Neumann entropy given by
\begin{equation}
S(\rho) = \sum_i h(\lambda_i)
\end{equation}
where $\lambda_i$ are the eigenvalues of the state $\rho$ and
$h(x)=-x\log_2 x$. But how can we divide the total correlations into a
classical and a quantum part? This question was first answered by
Henderson and Vedral~\cite{henderson2001classical} and Ollivier and
Zurek~\cite{ollivier2001quantum}. They defined the classical
correlations (CC) by
\begin{equation}
\label{eq_cc}
J(B|A)=S(B)-\min_{\{\Pi_a\}} \sum_a p_a S(\rho_{B|a}).
\end{equation}
where the $\{\Pi_a\}$ is a positive operator valued measure (POVM) performed on subsystem $A$. We refer to this measure of classical correlations as the asymmetric CC. A POVM describes a quantum measurement. It is set of nonnegative self-adjoint operators that satisfy $\sum_a \Pi_a = I$.  The probability of measuring outcome $a$ is $p_a=\Tr{\rho_A \Pi_a}$. The state of $B$ after $a$ is measured on $A$ is given by
\begin{equation}
\label{eq_rhobgivena}
\rho_{B|a}= \Tr_A\left(\frac{(M_a \otimes I) \rho_{AB} (M_a^\dagger \otimes I)}{p_a}\right),
\end{equation}
where $\Pi_a=M_aM_a^\dagger$. But what about the quantum correlations? They defined the quantum correlations or quantum discord (QD) as the total correlations minus the classical correlations,
\begin{equation}
\label{eq_discord}
\delta(B|A) = I_Q(A;B) - J(B|A).
\end{equation}
Henderson and Vedral defined classical correlations in this way because it satisfied certain desirable properties, and Ollivier and Zurek came up with this definition by generalizing classical conditional entropy to a quantum version. This is not the only way classical correlations can be defined.

One of the properties of the assymetric QD defined in
Eq.~(\ref{eq_discord}) is that it is not symmetric. That is,
$\delta(A|B) \neq \delta(B|A)$ in general. This is because the
asymmetric CC defined in Eq.~(\ref{eq_cc}) are not symmetric. However,
one desirable property of a measure of classical correlations $C$
would be that the measure is symmetric. This view was expressed by
Henderson and Vedral in their original
paper~\cite{henderson2001classical}: ``It is also natural that the
measure $C$ should be symmetric under interchange of the subsystems
$A$ and $B$. This is because it should quantify the correlation
between subsystems rather than a property of either subsystem." It was
not clear back then if the measure defined by Eq.~(\ref{eq_cc}) was
symmetric or not.

\subsection{Symmetric versions of quantum discord}
There are several different ways of defining a symmetric version of
the quantum discord, which turn out to be equivalent. We use the term
symmetric QD when Alice and Bob are restricted to performing projective measurements,
and extended symmetric QD when they can perform arbitrary POVM measurements.

\begin{figure*}
\includegraphics[width=18cm]{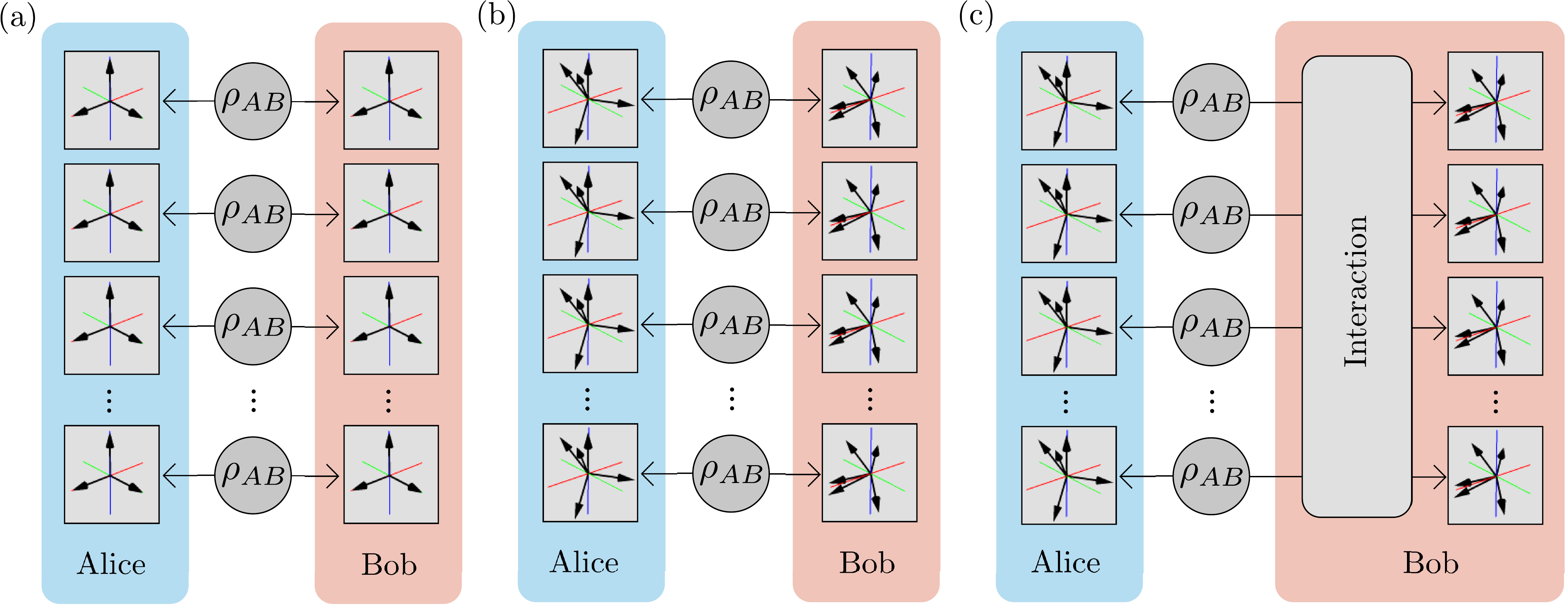}
\caption{ This diagram shows the different definitions of classical correlations (CC) of a bipartite state $\rho_{AB}$, that Alice and Bob share many copies of. (a) Alice and Bob perform projective measurements. The maximum classical mutual information (MI) of their measurement outcomes is the symmetric CC. (b) Alice and Bob perform POVM measurements. The maximum classical MI of their measurement outcomes is the extended symmetric CC. (c) Alice and Bob perform POVM measurements, but Bob is allowed to interact all his states beforehand. This is equivalent to the asymmetric CC. }
\label{fig_correlations}
\end{figure*}

\subsubsection{Symmetric quantum discord}

We first consider the approach of Ref.~\cite{maziero2010symmetry}, which was also used by Ref.~\cite{rulli2011global} to define the multipartite global quantum discord. 

Equation~(\ref{eq_cc}) can be written in an alternative form. Suppose Alice performs a projective measurement on her subsystem of $\rho_{AB}$. A projective measurement is a POVM in which the elements are $\Pi_a=\ket{a}\bra{a}$ where $\{\ket{a}\}$ are a set of states that form an orthonormal basis. The state after the measurement will be a classical-quantum state given by 
\begin{equation}
\label{eq_phia_rhoab}
\phi_A(\rho_{AB}) = \sum_a p_a \ket{a}\bra{a} \otimes \rho_{B|a}.
\end{equation}
Here we have defined $\phi_A(\rho)$ to be the state $\rho$ becomes after a measurement on $A$. After the measurement, $A$ will be diagonal in the measurement basis:
\begin{equation}
\phi_A(\rho_A) = \sum_a p_a \ket{a}\bra{a},
\end{equation}
with entropy $S(\phi_A(\rho_A))=\sum_a h(p_a)$. $B$ will be unchanged by a measurement on $A$: $\phi_A(\rho_B)=\rho_B$. Since $\phi_A(\rho_{AB})$ is a classical-quantum state, we can write its entropy as
\begin{equation}
\label{eq_ent_phia_rhoab}
S(\phi_A(\rho_{AB})) = S(\phi_A(\rho_A)) + \sum_a p_a S(\rho_{B|a}).
\end{equation}
The asymmetric CC Eq.~(\ref{eq_cc}), with the optimisation over
projective measurements instead of POVMs, is equivalent to the quantum
MI of state $\phi_A(\rho_{AB})$ maximized over all projective
measurements on $A$.
\begin{align}
\label{eq_cc_v2}
&\max_{\{\ket{a}\} } I_Q(\phi_A(\rho_{AB})) \\
&=\max_{\{\ket{a}\} } S(\phi_A(\rho_A))+S(\phi_A(\rho_B))-S(\phi_A(\rho_{AB})) \\
&=\max_{\{\ket{a}\} } S(\phi_A(\rho_A)) + S(\rho_B) - S(\phi_A(\rho_A)) -\sum_a p_a S(\rho_{B|a}) \\
\label{eq_cc_v2_last}
&= S(B) - \min_{\{\ket{a}\}}\sum_a p_a S(\rho_{B|a})
\end{align}
The interpretation of the measurement that maximizes the above is that it is the projective measurement that least disturbs the state, that is, the projective measurement that results in the least loss in quantum MI. 

The extension to a symmetric version is simple. In this case, a projective measurement is performed on both $A$ and $B$. After the measurement, $\rho_{AB}$ will became be a classical-classical state given by
\begin{equation}
\label{eq_phiAB_rhoAB}
\phi_{AB}(\rho_{AB}) = \sum_{a,b} p_{ab} \ket{a}\bra{a} \otimes \ket{b}\bra{b}.
\end{equation}
where $p_{ab}=\Tr{\rho_{AB} \ket{a}\bra{a}\otimes \ket{b}\bra{b}}$. The symmetric CC is given by
\begin{align}
\label{eq_js}
J_S(A;B)&=\max_{\{\ket{a}\},\{\ket{b}\}} I_Q(\phi_{AB}(\rho_{AB})) \\
&=\max_{\{\ket{a}\},\{\ket{b}\}} S(\phi(\rho_A))+S(\phi(\rho_B))-S(\phi(\rho_{AB})) \label{eq_js2} \\
&=\max_{\{\ket{a}\},\{\ket{b}\}} \sum_a h(p_a) + \sum_b h(p_b) -\sum_{a,b} h(p_{ab}) \label{eq_js3}
\end{align}
The symmetric QD is
\begin{equation}
\delta_S(A;B) = I_Q(A;B) - J_{S}(A;B).
\end{equation}
The interpretation of the symmetric QD is that it is the smallest loss in quantum MI after local projective measurements on $A$ and $B$. 

The asymmetric CC and symmetric CC are related by the inequality
\begin{equation}
\label{eq_discord_ineq}
J_S(A;B) \le J(A|B).
\end{equation}
This inequality results from the fact that the quantum mutual information of a state cannot increase under a measurement of one subsystem. Hence, the correlations after measurement reduce (or remain the same) if Bob does a measurement in addition to Alice.

\subsubsection{Extended symmetric quantum discord}
\label{sec_esqd}

An alternative way of defining the classical correlations is using the classical mutual information~\cite{piani2008no, wu2009correlations, terhal2002entanglement}. This turns out to be equivalent to symmetric QD but is defined when the measurement is any POVM.

Let $\mathcal{A}$ be the random variable that describes measurement outcomes $a$ on state $A$, and $\mathcal{B}$ be the random variable that describes measurement outcomes $b$ on state $B$. The classical MI between $\mathcal{A}$ and $\mathcal{B}$ is
\begin{equation}
I_C(\mathcal{A},\mathcal{B}) =H(\mathcal{A}) + H(\mathcal{B}) - H(\mathcal{AB}),
\end{equation}
where $H(X)$ is the Shannon entropy of variable $X$. It is defined by
\begin{equation}
H(X) = \sum_x h(p_x),
\end{equation}
where $p_x$ is the probability that $X=x$.

We introduce a new measure of the classical correlations $J_{ES}$, we call the extended symmetric CC. It is defined as the maximum classical mutual information between the measurement outcomes made on $A$ and $B$, maximized over all POVM measurements. 
\begin{align}
\label{eq_cs}
J_{ES}(A;B)&=\max_{\{\Pi_a\},\{\Pi_b\}} I_C(\mathcal{A},\mathcal{B}) \\
&=\max_{\{\Pi_a\},\{\Pi_b\}} \sum_a h(p_a) + \sum_b h(p_b) -\sum_{a,b} h(p_{ab}) \label{eq_cs2} \\
&=\max_{\{\Pi_a\},\{\Pi_b\}} \sum_b h(p_b) - \sum_a p_a \sum_b h(p_{b|a}) \label{eq_cs3}
\end{align}
This quantity is symmetric, because classical mutual information is symmetric. We then define the extended symmetric QD as
\begin{equation}
\delta_{ES}(A;B) = I_Q(A;B) - J_{ES}(A;B).
\end{equation}

The same quantity is being maximised in Eqs.~(\ref{eq_js3}) and
(\ref{eq_cs2}). This implies two things. Firstly, there is an
equivalent interpretation of the symmetric CC as the classical MI
between measurement outcomes on $A$ and $B$, maximised over local
projective measurements. Secondly, we have the inequality,
\begin{equation}
\label{eq_symineq}
J_{ES}(A;B) \ge J_S(A;B),
\end{equation}
because projective measurements are a subset of POVMs. The extended symmetric CC can be viewed as an extension of the symmetric CC, in that it extends the definition of symmetric CC to general POVMs. 

\subsection{Comparison of CC measures}

Figure \ref{fig_correlations} shows the interpretations of the three
different CC quantities. Suppose Alice and Bob share $n$ copies of a
bipartite state $\rho_{AB}$. Alice measures each copy separately using
the same measurement. Let $\mathcal{A}$ be the random variable that
describes her measurement outcomes. Bob does the same thing, and
$\mathcal{B}$ describes his measurement outcomes. The maximum
classical MI between $\mathcal{A}$ and $\mathcal{B}$ is the symmetric
CC if Alice and Bob can perform projective measurements, extended CC
if they can do any POVM measurements.

Now suppose Bob is allowed to interact all his copies before he measures
them, as in Figure \ref{fig_correlations}(c). Can he gain any more
information about Alice's measurements outcomes? The answer is yes,
provided Alice sends Bob some additional classical information. The
maximum information Bob can obtain about Alice's measurement outcome subtracting the
additional classical information Alice sends is equal to the
asymmetric CC. A protocol that achieves this rate is described in Ref.~\cite{devetak2003classical}.

These quantities are related by
\begin{equation}
\label{eq_ccineq}
\min \left(J(A|B),J(B|A) \right) \ge J_{ES}(A;B) \ge J_S(A;B)\;.
\end{equation}

\subsection{Multipartite quantum discord}
\label{sec_multi_sqd}

The extended symmetric CC and QD can be defined for multipartite
states~\cite{piani2008no}, using multipartite extensions of the
classical MI and quantum MI~\cite{watanabe1960information}. Let a
multipartite state $\rho_{\vec{A}}$ be distributed to $n$ parties, where $\vec{A}=[A_1,A_2,\ldots,A_n]$. Let
$A_i$ denote the subsystem received by $i$-th party. Each party
measures their subsystem, and $\mathcal{A}_i$ denotes the random
variable that describes measurement outcomes $a_i$ on subsystem
$A_i$. The multipartite classical MI is
\begin{equation}
I_C(\mathcal{A}_1;\mathcal{A}_2;\ldots;\mathcal{A}_n) = \sum_{i=1}^n
H(\mathcal{A}_i) - H(\mathcal{A}_1 \mathcal{A}_2 \ldots  \mathcal{A}_n),
\end{equation}
The $n$-partite extended symmetric CC of state $\rho_{\vec{A}}$ is 
\begin{align}
\label{eq_mescc}
&J_{ES}(A_1;A_2;\ldots;A_n) \nonumber \\
=& \max_{\{\Pi_{a_1}\},\{\Pi_{a_2}\},\ldots,\{\Pi_{a_n}\}} I_C(\mathcal{A}_1;\mathcal{A}_2;\ldots;\mathcal{A}_n),
\end{align}
The maximization is over local POVM measurements $\{\Pi_{a_i}\}$ performed on subsystems $A_i$.
\begin{equation}
H(\mathcal{A}_i) = \sum_{a_i} h(p_{a_i}),
\end{equation}
where $p_{a_i}=\Tr{\rho_{A_i}\Pi_{a_i}}$.
\begin{equation}
H(\vec{\mathcal{A}}) = \sum_{a_1}\sum_{a_2}\ldots\sum_{a_n} h(p(a_1,a_2,\ldots,a_n)),
\end{equation}
where $p(a_1,a_2,\ldots,a_n)=\Tr{\rho_{\vec{A}}\Pi_{a_1}\Pi_{a_2}\ldots\Pi_{a_n}}$.

Similarly, the multipartite quantum MI is given by
\begin{equation}
I_Q(A_1;A_2;\ldots;A_n) = \sum_{i=1}^n S(A_i) - S(A_1 A_2 \ldots A_n).
\end{equation}
This allows us to define the multipartite extended symmetric QD by 
\begin{multline}
\label{eq_mesqd}
\delta_{ES}(A_1;A_2;\ldots;A_n) = I_Q(A_1;A_2;\ldots;A_n) \\
 - J_{ES}(A_1;A_2;\ldots;A_n).
\end{multline}
Rulli and Sarandy defined a similar quantity called the global
quantum discord~\cite{rulli2011global}, but the measurements
are restricted to projective measurements. The multipartite extended symmetric QD can be viewed as an extension
of the global quantum discord to general POVM measurements.

\section{Gaussian multipartite CC and QD}
\label{sec_gaussian_multi}

\begin{figure}
\includegraphics[width=8.0cm]{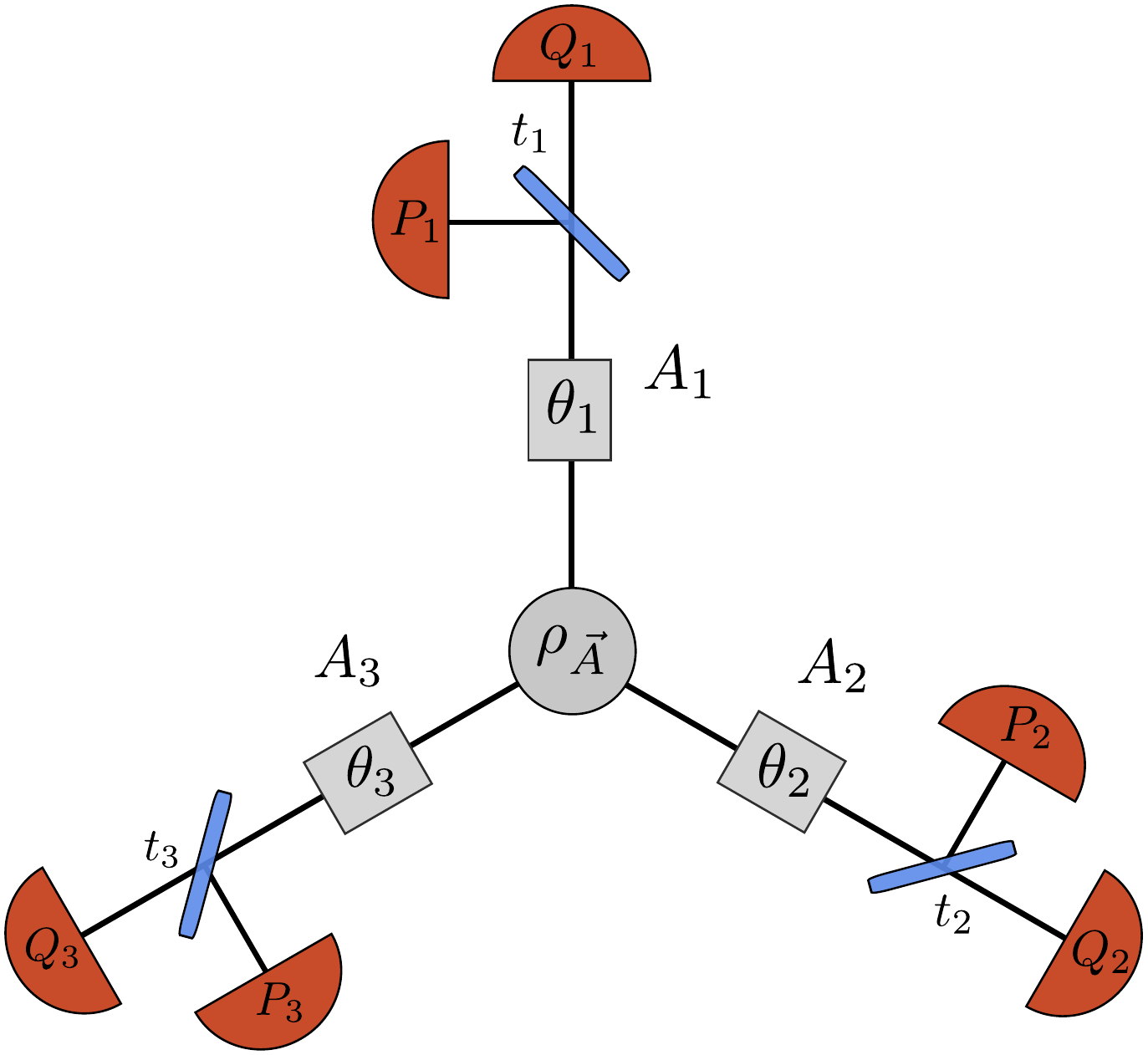}
\caption{ The three modes of a tripartite Gaussian state $\rho_{\vec{A}}$ are distributed to three parties. Each party does a Gaussian measurement on their subsystem $A_i$. A Gaussian measurement of a single-mode Gaussian state consists of a phase shift of $\theta_i$ followed by a beam splitter with transmissivity $0\le t_i \le 1$ followed by measurement of the $Q$ and $P$ quadratures. The maximum classical mutual information between the measurement outcomes gives the multipartite Gaussian CC (Eq.~(\ref{eq_cg_n})). }
\label{fig_measure}
\end{figure}

Let us define the Gaussian multipartite CC $J_G(\vec{A})$ to be the maximum classical MI achievable when the measurement on each subsystem are restricted to Gaussian measurements. Hence $J_G(\vec{A})$ is equivalent to Eq.~(\ref{eq_mescc}) except the maximization is over Gaussian POVMs, rather than all POVMs.

We introduce the Gaussian multipartite QD given by 
\begin{equation}
\label{eq_cs_g}
\delta_G(\vec{A}) = I_Q(\vec{A}) - J_G(\vec{A}).
\end{equation}
Suppose $n$ parties each receive one mode of an $n$-partite Gaussian state. We now describe how to calculate the Gaussian multipartite QD in this situation. Each party performs a Gaussian measurement on their subsystem. A Gaussian measurement of a single-mode Gaussian state can be described by a phase shift $\theta_i$ followed by a beam splitter with transmissivity $0\le t_i\le 1$ and orthogonal quadrature measurements $Q_i$ and $P_i$ on the outputs of the beam splitter. Figure \ref{fig_measure} shows a diagram of the measurements performed for the case of a tripartite Gaussian state.

The Gaussian multipartite CC of a bipartite state is
\begin{multline}
\label{eq_cg_0}
J_G(A_1,A_2) = \max_{\theta_1,t_1,\theta_2,t_2} I_C(Q_2;Q_1P_1)+I_C(P_2|Q_2;Q_1P_1) \\
= \max_{\theta_1,t_1,\theta_2,t_2} \big[G(Q_2) - G(Q_2|Q_1P_1) + G(P_2|Q_2)
\\
- G(P_2|Q_1P_1Q_2)\big]\;,
\end{multline}
where $G(Q)$ is the differential entropy of $Q$.
The measurement outcome of a Gaussian measurement performed on a
Gaussian state are normally distributed. Consider a random variable
$X$ that is normally distributed. The probability density of $X$ is
\begin{equation}
\label{eq_px}
p_x = \frac{1}{\sqrt{2\pi V_X}} \exp(-\frac{(x-\mu_X)^2}{2V_X}),
\end{equation}
where $\mu_X$ is the mean of $X$ and $V_X$ is the variance of $X$. The differential entropy of $X$ is
\begin{align}
\label{eq_HX1}
G(V_X) &= -\int_{-\infty}^{\infty} p_x \log_2 p_x \;dx\\
\label{eq_HX2} &= \frac{1}{2} \log_2 (2 \pi e V_X),
\end{align}
Note that the differential entropy of $X$ does not depend on the mean of $X$.

There is an additional property that allows us to simplify
Eq.~(\ref{eq_cg_0}). The conditional variances do not depend on other
measurement outcomes. For example, the variance of $Q_2$ conditioned
on the measurement outcomes of $Q_1$ and $P_1$, denoted
$V_{Q_2|Q_1 P_1}$, will be a constant independent of the measurement
outcomes of $Q_1$ and $P_1$. Therefore, Eq.~(\ref{eq_cg_0}) becomes
\begin{multline}
\label{eq_cg_n2}
J_G(A_1,A_2) = \max_{\theta_1,t_1,\theta_2,t_2} \Big[ G(V_{Q_2}) \\
+ G(V_{P_2|Q_2}) - G(V_{Q_2|Q_1 P_1}) - G(V_{P_2|Q_1 P_1 Q_2}) \Big].
\end{multline}
The extension to $n$-partite states is
\begin{align}
\label{eq_cg_n}
J_G(\vec{A}) =&\ \max_{\theta_1,t_1,\theta_2,t_2,\ldots,\theta_n,t_n} \big[ \nonumber\\
&\ H(\mathcal{A}_2) - H(\mathcal{A}_2|\mathcal{A}_1) \nonumber\\
+&\ H(\mathcal{A}_3) - H(\mathcal{A}_3|\mathcal{A}_2 \mathcal{A}_1) \nonumber\\
\vdots \nonumber\\
+&\ H(\mathcal{A}_n) - H(\mathcal{A}_n|\mathcal{A}_{n-1}
   \mathcal{A}_{n-2} \ldots A_1)
\big],
\end{align}
where
\begin{equation}
H(\mathcal{A}_i) = G(V_{Q_i}) + G(V_{P_i|Q_i})
\end{equation}
and 
\begin{multline}
H(\mathcal{A}_i|\mathcal{A}_{i-1} \ldots \mathcal{A}_1) =
G(V_{Q_i|Q_{i-1} P_{i-1} \ldots Q_1 P_1}) \\
+G(V_{P_i|Q_i Q_{i-1} P_{i-1} \ldots Q_1 P_1}).
\end{multline}
The calculation of Eq.~(\ref{eq_cg_n}) involves an optimization over $2n$ variables. We will demonstrate the calculation of Eq.~(\ref{eq_cg_n}) for some states. 

\subsection{Properties}

We state and prove several desirable properties of the the Gaussian multipartite CC and QD.

\begin{enumerate}
\item Gaussian multipartite CC is symmetric. This is true because the classical mutual information is symmetric.
\item The Gaussian multipartite QD is zero for product states. This follows from the nonnegativity of the Gaussian multipartitie QD and the fact that quantum mutual information is zero for product states.
\item The Gaussian multipartite CC does not increase under local Gaussian operations. This is because local Gaussian operations can be considered part of the measurements.
\end{enumerate}

\subsection{Example: two-mode EPR state}

\begin{figure*}
\includegraphics[width=17cm]{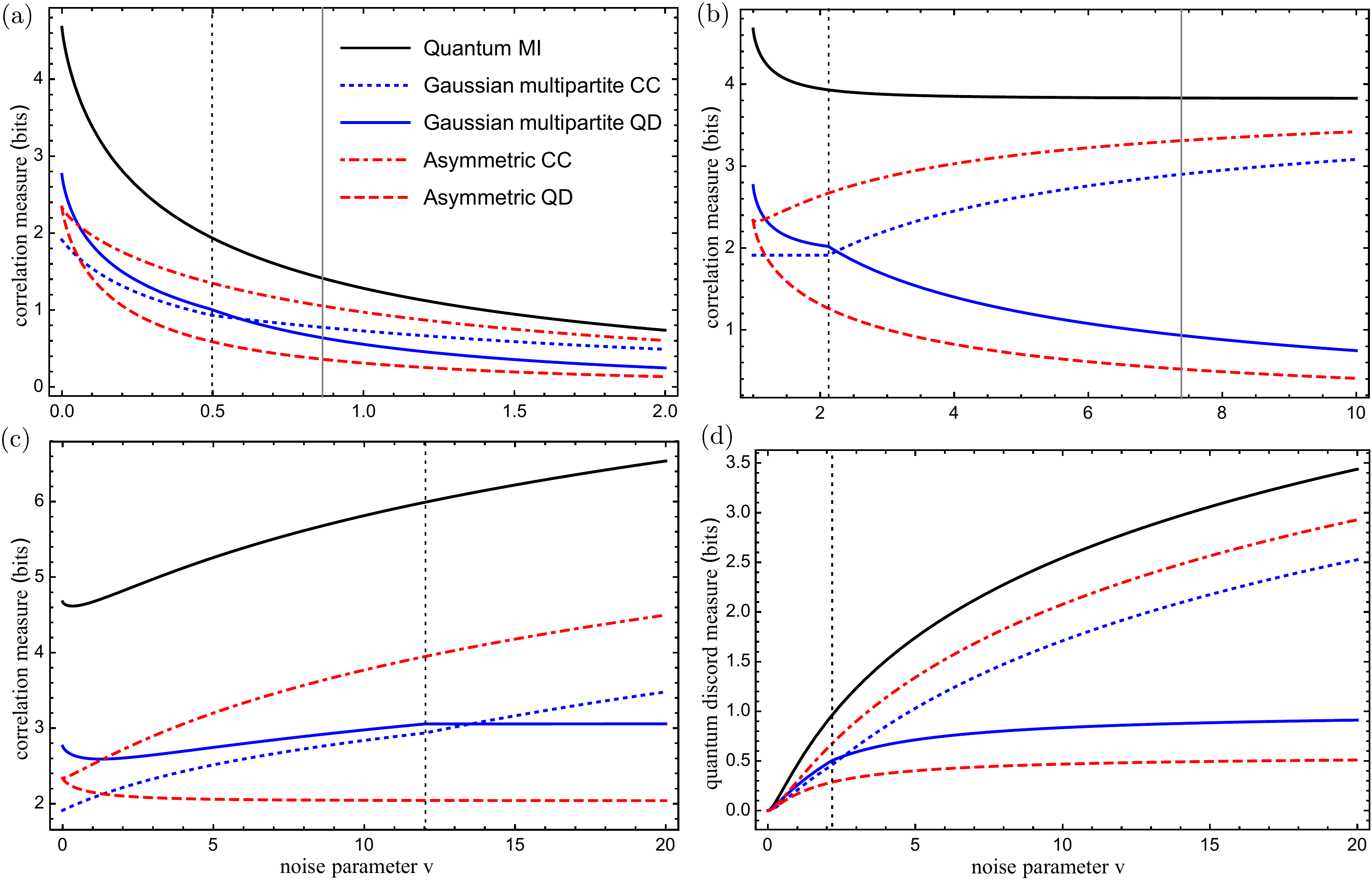}
\caption{ This plot shows the correlations present in a
  two-mode squeezed EPR state (Eq.~(\ref{eq:EPR})) with $r=1$ subjected to (a) uncorrelated noise, (b) multiplicative noise, and (c) correlated noise. Also shown is (d) the correlations present in a vacuum state subjected to correlated noise. If $v$ is less than the dashed vertical line, the measurements that attains the Gaussian multipartite QD and CC are homodyne measurements, otherwise the measurements are heterodyne measurements. For $v$ less than the solid vertical line the state is entangled, otherwise separable. For (c), the state is entangled for all $v$. For (d), the state is separable for all $v$.}
\label{fig_plot1}
\end{figure*}

Consider a two-mode Einstein-Podolsky-Rosen (EPR) state. The covariance of the quadratures of the EPR state is~\cite{weedbrook2012gaussian}
\begin{equation}
  \label{eq:EPR}
V_{EPR} = \begin{pmatrix}
\cosh 2r & 0 & \sinh 2r & 0 \\
0 & \cosh 2r & 0 & -\sinh 2r \\
\sinh 2r & 0 & \cosh 2r & 0 \\
0 & -\sinh 2r & 0 & \cosh 2r
\end{pmatrix}.
\end{equation}
The measurement that attains the Gaussian multipartite CC will have
phase shifts of zero, i.e.\ $\theta_1=\theta_2=0$. In fact, the phase
shifts will be zero for any quadrature covariance matrix that has zero
covariance between $Q$ and $P$ quadratures. Values of the other
parameters $t_1$ and $t_2$ were found by performing the optimisation
analytically. The optimum occurs when $t_1=t_2=0$ or $1$, giving
$J_G(A_1,A_2)=\log_2(\cosh 2r)$. This corresponds to performing
homodyne measurements on each subsystem.

If $r=1$ then $J_G = 1.912$. If instead we consider the extended
symmetric CC, where the measurements are not restricted to Gaussian
measurements we obtain $J_{ES} = 2.337$. This is obtained when both parties measure
in the Fock number state basis. By
restricting to Gaussian measurements we reduce the amount of classical
correlations that can be seen. The calculation of Gaussian classical
correlations however, is much simpler. In general, it is nontrivial to
find the measurement that optimises $J_{ES}$. Additionally, Gaussian measurements have the
added bonus of being easy to do experimentally, requiring only linear
optical elements and homodyne measurements.

\subsection{Example: noisy EPR state}

We calculated the multipartite Gaussian CC and QD for an EPR state subjected to three different types of noise, which is plotted in Fig.\ \ref{fig_plot1}(a,b,c). A useful result, derived by Ref.\ \cite{mivsta2016gaussian} for the calculation of Gaussian intrinsic entanglement, is that for a state with quadrature covariance matrix
\begin{equation}
\begin{pmatrix}
a & 0 & c_x & 0 \\
0 & a & 0 & c_p \\
c_x & 0 & b & 0 \\
0 & c_p & 0 & b
\end{pmatrix}
\end{equation}
with $c_x\ge|c_p|\ge 0$, the Gaussian multipartite CC of this state is obtained by a homodyne measurements of the $Q$ quadratures (corresponding to the measurement when $t_1=t_2=1$) if
\begin{equation}
\sqrt{\frac{a}{b}} + \sqrt{\frac{b}{a}} + \frac{1}{\sqrt{{a}{b}}}-\sqrt{ab-c_x^2} \ge 0.
\end{equation}
Since the two-mode states we consider are symmetric in the $Q$ and $P$ quadratures, homodyne measurements of the $P$ quadratures (corresponding to the measurement when $t_1=t_2=0$) gives the same classical MI. When the above inequality is not satisfied, numerical optimisation revealed that for all two mode states we considered the optimal measurement is a heterodyne measurement of both modes (corresponding to $t_1=t_2=1/2$).

As is typical of quantum discord quantities, we observe that when noise is increased sufficiently such that the state becomes separable, determined using Duan's inseparability criterion~\cite{duan2000inseparability}, there is still a nonzero amount of Gaussian multipartite QD. 

We also calculate the asymmetric CC and QD. For the states we
consider, the asymmetric QD is equal to the asymmetric Gaussian QD,
and additionally this is obtained by a heterodyne measurement on one
of the subsystems~\cite{pirandola2014optimality}. We are unaware of
any simple means of calculating the symmetric QD or extended symmetric
QD for Gaussian states, so we chose not to calculate these quantities.

\subsubsection{Uncorrelated noise}

Firstly let us consider the case in which uncorrelated quadrature noise is added to each mode of the EPR state. The quadrature covariance matrix of the resulting state is $V_{EPR}+v I_4$ where $I_4$ is the 4-by-4 identity matrix, and $v\ge0$ is a parameter that controls the amount of noise. A plot of correlation is shown in Fig.~\ref{fig_plot1}(a). The total correlations, as measured by the quantum MI, decreases as $v$ increases. The Gaussian multipartite CC, Gaussian multipartite QD, assymetric CC, and symmetric CC also all decrease as the noise increases. 

\subsubsection{Multiplicative noise}

Now consider the case in which the quadrature covariance is multiplied by a factor $v\ge1$, so the quadrature covariance matrix is $v V_{EPR}$. This type of noise is realised if the EPR state is generated by mixing on a beam splitter two squeezed states that are impure. Then $v$ is equal to multiplication of the squeezed state quadrature variances. The information quantities are shown in Fig.~\ref{fig_plot1}(b). The total correlations, as measured by the quantum MI, decreases as $v$ increases. Despite this, the Gaussian multipartite CC and asymmetric CC do increase, however this is at the expense of the Gaussian multipartite QD and asymmetric QD, which decrease. 

For $v$ less than some value, the Gaussian multipartite CC is constant. This is the region in which a homodyne measurement is optimal.

\subsubsection{Correlated noise}

The third case we consider is adding classically correlated noise to $Q$ quadratures of each mode, and classically anticorrelated noise to the $P$ quadratures. The quadrature covariances of the resulting state is 
\begin{equation}
V_{EPR}+
\begin{pmatrix}
v & 0 & v & 0 \\
0 & v & 0 & -v \\
v & 0 & v & 0 \\
0 & -v & 0 & v
\end{pmatrix}. 
\end{equation}
The information quantities are shown in Fig.~\ref{fig_plot1}(c). Unsurprisingly, the Gaussian multipartite CC and the asymmetric CC increase as a function of $v$, because we are adding classically correlated noise. 

Adding correlated noise initially reduces the asymmetric QD and Gaussian multipartite QD, which also results in a dip in the quantum MI at the start. The heterodyne measurement is much better for detecting the added classical correlations, so when the heterodyne measurement is optimal the Gaussian multipartite QD is almost constant. For large $v$, the asymmetric QD and Gaussian multipartite QD appear almost constant but they are in fact slowly increasing.

It is perhaps counterintuitive that classically correlated noise can increase quantum discord. This can be more easily seen in Fig.~\ref{fig_plot1}(d), where classically correlated noise is added to a vacuum state. The state initially has zero correlations, but when the noise is added, all the correlation measures increase, including Gaussian multipartite QD and asymmetric QD. Generating a state with nonzero assymetric QD in this manner was done experimentally by \cite{gu2012observing}.

\subsection{Example: noisy Gaussian tripartite GHZ state}

\begin{figure*}
\includegraphics[width=17cm]{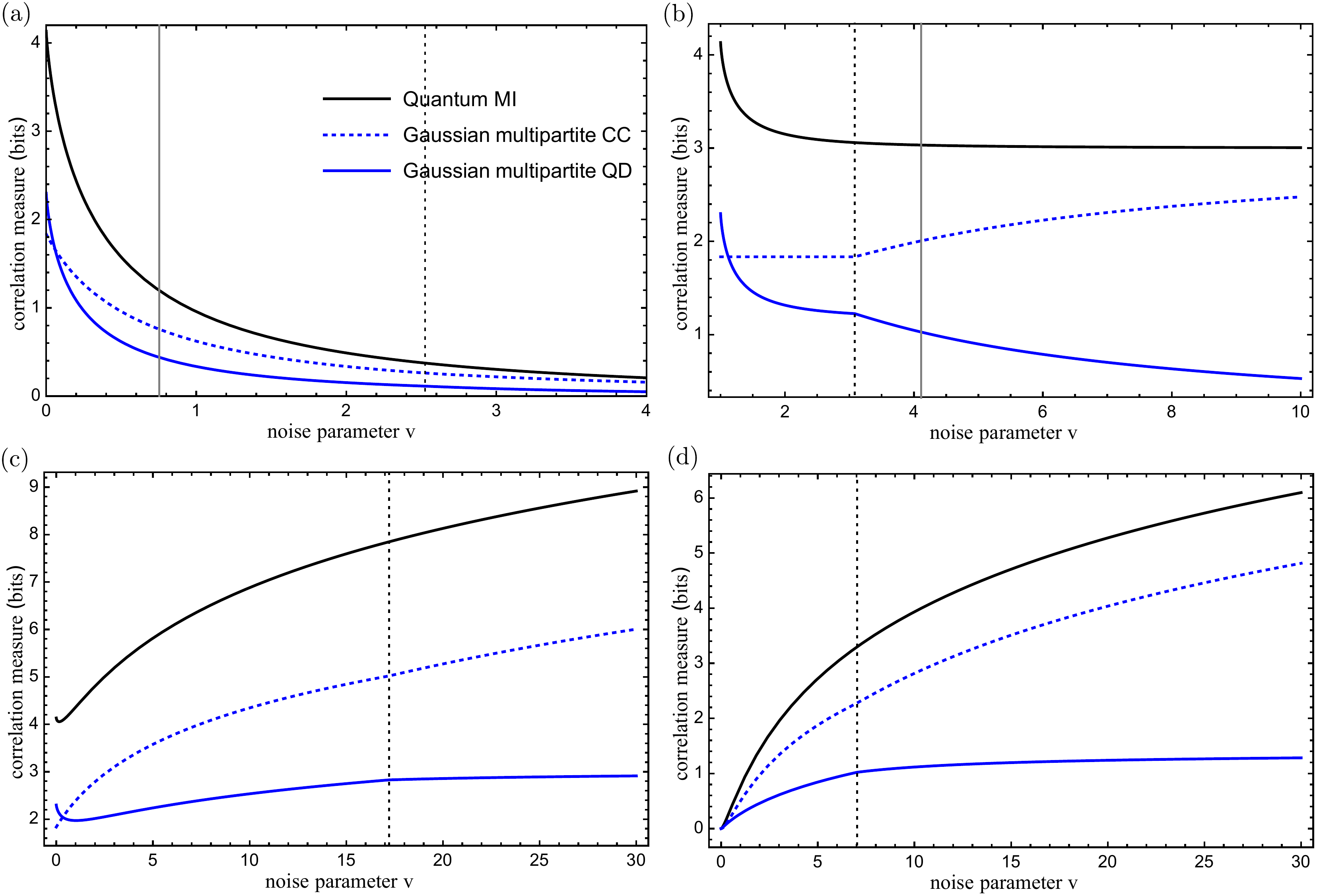}
\caption{ The correlations present in a
  three-mode Gaussian GHZ state (Eq.~(\ref{eq:GHZ})) with $a=2$ subjected to (a) uncorrelated noise, (b) multiplicative noise, (c) correlated noise, and (d) the correlations present in a three-mode vacuum state subjected to correlated noise. If $v$ is less than the dashed vertical line, the measurements that attains the Gaussian multipartite CC and QD are homodyne measurements.  For $v$ less than the solid vertical line the state is entangled, otherwise separable. For (c), the state is entangled for all $v$. For (d), the state is separable for all $v$. }
\label{fig_plot2}
\end{figure*}

The tripartite Gaussian state equivalent to the three-qubit GHZ and W states is a state with quadrature covariance matrix given by~\cite{adesso2006multipartite}
\begin{equation}
  \label{eq:GHZ}
V_{GHZ} =
\begin{pmatrix}
 a & 0 & c^+ & 0 & c^+ & 0 \\
 0 & a & 0 & c^- & 0 & c^- \\
 c^+ & 0 & a & 0 & c^+ & 0 \\
 0 & c^- & 0 & a & 0 & c^- \\
 c^+ & 0 & c^+ & 0 & a & 0 \\
 0 & c^- & 0 & c^- & 0 & a \\
\end{pmatrix}
\end{equation}
where 
\begin{equation}
c^\pm = \frac{a^2-1\pm \sqrt{(a^2-1)(9a^2-1)}}{4a}.
\end{equation}
Like for the two-mode case, we calculate the Gaussian multipartite CC and QD for the state subjected to three different types of noise. Figure \ref{fig_plot2} shows our results. To determine whether a state is separable, we use the method of ~\cite{giedke2001separability}.

\subsubsection{Uncorrelated noise}

Consider a three-mode GHZ state with uncorrelated quadrature noise added to each of the three modes. The resulting state has a quadrature covariances $V_{GHZ} + v I_6$ where $I_6$ is the 6-by-6 identity matrix. Figure \ref{fig_plot2}(a) is a plot of the information quantities. Just as in the two mode case, the quantum MI, Gaussian multipartite CC and QD all decrease as $v$ increases. 

\subsubsection{Multiplicative noise}

The information quantities for a three-mode GHZ state with multiplicative noise, i.e.\ a state with covariance matrix $v V_{GHZ}$, are shown in Fig.\ \ref{fig_plot2}(b). Similar to the two made case, this type of noise reduces the total correlations (quantum MI) and Gaussian multipartite QD, while at the same time increasing the Gaussian multipartite CC when $v$ is large. Just as in the two-mode case, homodyne measurements on each mode give the a classical mutual information that does not depend on $v$. Hence, the Gaussian multipartite CC is constant when the optimal measurement consists of homodyne measurements, which for $a=2$, is when $v<3.082$. 

\subsubsection{Correlated noise}

\begin{figure}
\includegraphics[width=8.5cm]{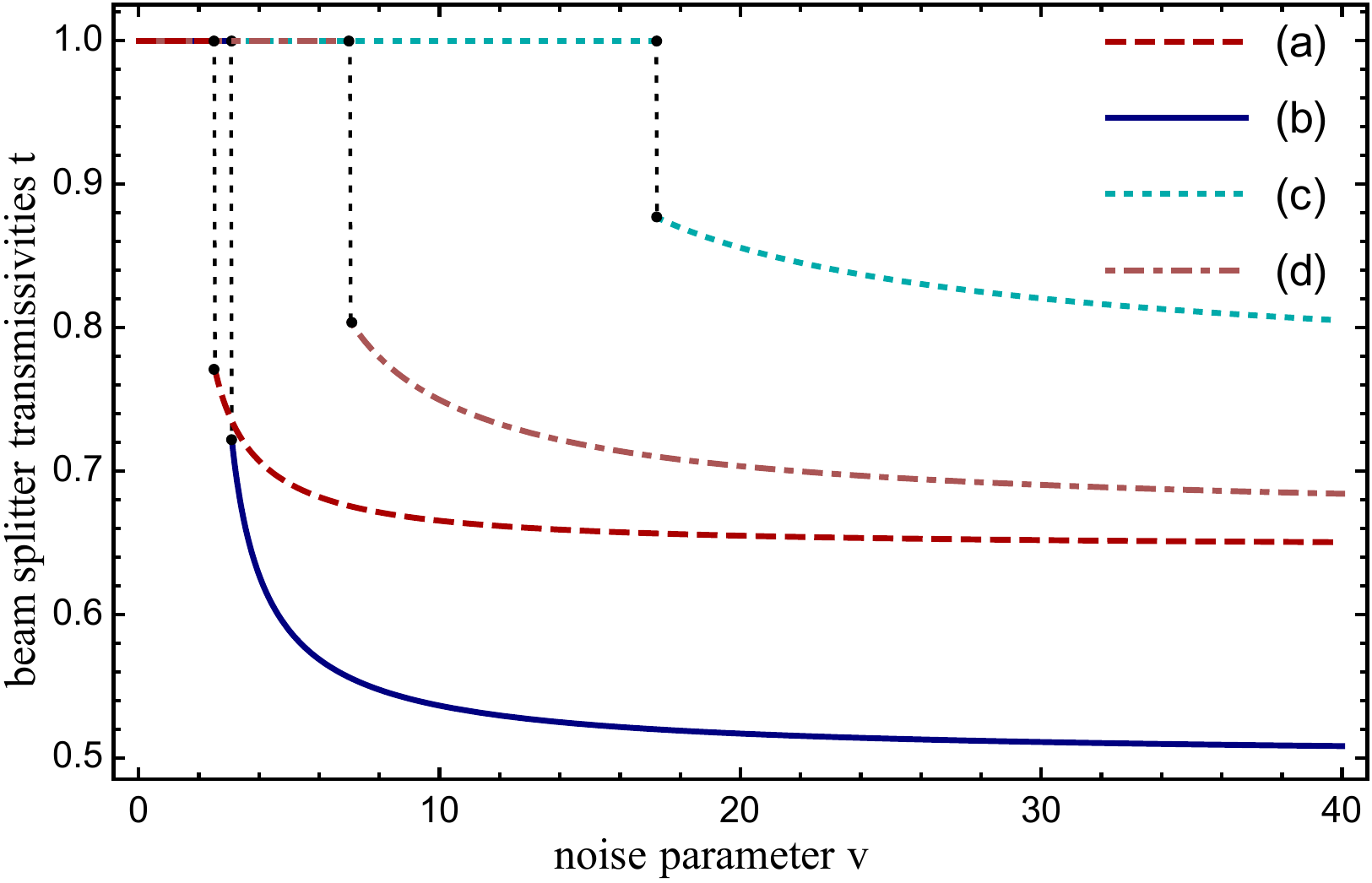}
\caption{This plot shows the beam splitter transmissivities $t_1=t_2=t_3=t$ for the measurement that attains the Gaussian multipartite CC and QD for the corresponding plots in Fig.\ \ref{fig_plot2}. }
\label{fig_plot3}
\end{figure}

Now we consider the case in which correlated noise is added to the $Q$ quadratures of each mode and anticorrelated noise is added to the $P$ quadratures. The resulting state has quadrature covariances
\begin{equation}
V_{GHZ} + v
\begin{pmatrix}
1 & 0 & 1 & 0 & 1 & 0 \\
0 & 1 & 0 & -0.5 & 0 & -0.5 \\
1 & 0 & 1 & 0 & 1 & 0 \\
0 & -0.5 & 0 & 1 & 0 & -0.5 \\
1 & 0 & 1 & 0 & 1 & 0 \\
0 & -0.5 & 0 & -0.5 & 0 & 1
\end{pmatrix}.
\end{equation}
Note that the matrix contains $-0.5$ terms. This is the largest anticorrelation that three classical variables with variance of 1 can have. 

The information quantities for this state are shown in Fig.\ \ref{fig_plot2}(c). We notice three properties that are the same as the two-mode case. (1) Initially there is a dip in the quantum MI and Gaussian multipartite QD. (2) The Gaussian mulitpartite CC increase as a function of $v$. (3) While homodyne measurements are optimal, after the initial dip, the multipartite QD increases as a function of $v$. When homodyne measurements are not optimal, the mulitpartite QD appears almost constant but is in fact slowly increasing. 

Figure \ref{fig_plot2}(d) shows the correlations present when the correlated noise is added to a three-mode vacuum state. Just as in the two-mode case, the Gaussian multipartite QD discord is nonzero for $v\ge0$, despite the fact that the state is separable. 

\subsubsection{Measurements}

In all of the cases described above, for small $v$, the measurement that attains the Gaussian multipartite CC and QD consists of homodyne measurements of the $Q$ quadrature on each mode ($t_1=t_2=t_3=1$). Homodyne measurements of the $P$ quadratures does not give the same classical MI because there is an asymmetry in the $Q$ and $P$ quadratures of the Gaussian GHZ state; the anticorrelations of the $P$ quadratures are less than the correlations of the $Q$ quadratures.

For large $v$, the optimal measurement consists of beam splitter transmisivities $t_1=t_2=t_3=t$ where $t<1$. In stark contrast to the two-mode case, this value of $t$ depends on $v$. A plot the relationship between $v$ and $t$ is shown in Fig.\ \ref{fig_plot3}. There is a discontinuity at the point where homodyne measurements are no longer optimal; the value of $t$ abruptly changes from 1 to some value that is less than 1. Note that there is also a discontinuity in the two made case, in which $t$ changes from 1 or 0 to $1/2$. 

\section{Conclusion}
\label{sec_conclusion}

We have introduced a new measure of the classical correlations of a
multipartite Gaussian state, defined as the maximum classical MI
between Gaussian measurement outcomes performed on each subsystem. We
introduce a new measure of multipartite Gaussian QD defined by
subtracting the multipartite Gaussian CC from the multiparite quantum
MI of the state. The Gaussian multipartite CC is easy to calculate,
requiring an optimisation over at most $2n$ variables for an $n$-mode
Gaussian state. We envisage this measure being relevant in Gaussian
quantum information experiments that do not use any non-Gaussian measurements.

We calculated the Gaussian multipartite CC and QD for a two-mode EPR state and a three-mode Gaussian GHZ state subjected to different types of noise. 

\section*{Acknowledgements} 
This research is supported by the
Australian Research Council (ARC) under the Centre of Excellence for
Quantum Computation and Communication Technology (CE110001027). We acknowledge funding from 
the Defence Science and Technology group. We would like to thank Mile Gu for discussions on the paper.

\bibliography{bibliography}

\begin{thebibliography}{27}%
\makeatletter
\providecommand \@ifxundefined [1]{%
 \@ifx{#1\undefined}
}%
\providecommand \@ifnum [1]{%
 \ifnum #1\expandafter \@firstoftwo
 \else \expandafter \@secondoftwo
 \fi
}%
\providecommand \@ifx [1]{%
 \ifx #1\expandafter \@firstoftwo
 \else \expandafter \@secondoftwo
 \fi
}%
\providecommand \natexlab [1]{#1}%
\providecommand \enquote  [1]{``#1''}%
\providecommand \bibnamefont  [1]{#1}%
\providecommand \bibfnamefont [1]{#1}%
\providecommand \citenamefont [1]{#1}%
\providecommand \href@noop [0]{\@secondoftwo}%
\providecommand \href [0]{\begingroup \@sanitize@url \@href}%
\providecommand \@href[1]{\@@startlink{#1}\@@href}%
\providecommand \@@href[1]{\endgroup#1\@@endlink}%
\providecommand \@sanitize@url [0]{\catcode `\\12\catcode `\$12\catcode
  `\&12\catcode `\#12\catcode `\^12\catcode `\_12\catcode `\%12\relax}%
\providecommand \@@startlink[1]{}%
\providecommand \@@endlink[0]{}%
\providecommand \url  [0]{\begingroup\@sanitize@url \@url }%
\providecommand \@url [1]{\endgroup\@href {#1}{\urlprefix }}%
\providecommand \urlprefix  [0]{URL }%
\providecommand \Eprint [0]{\href }%
\providecommand \doibase [0]{http://dx.doi.org/}%
\providecommand \selectlanguage [0]{\@gobble}%
\providecommand \bibinfo  [0]{\@secondoftwo}%
\providecommand \bibfield  [0]{\@secondoftwo}%
\providecommand \translation [1]{[#1]}%
\providecommand \BibitemOpen [0]{}%
\providecommand \bibitemStop [0]{}%
\providecommand \bibitemNoStop [0]{.\EOS\space}%
\providecommand \EOS [0]{\spacefactor3000\relax}%
\providecommand \BibitemShut  [1]{\csname bibitem#1\endcsname}%
\let\auto@bib@innerbib\@empty
%</preamble>
\bibitem [{\citenamefont {Horodecki}\ \emph {et~al.}(2009)\citenamefont
  {Horodecki}, \citenamefont {Horodecki}, \citenamefont {Horodecki},\ and\
  \citenamefont {Horodecki}}]{horodecki2009quantum}%
  \BibitemOpen
  \bibfield  {author} {\bibinfo {author} {\bibfnamefont {R.}~\bibnamefont
  {Horodecki}}, \bibinfo {author} {\bibfnamefont {P.}~\bibnamefont
  {Horodecki}}, \bibinfo {author} {\bibfnamefont {M.}~\bibnamefont
  {Horodecki}}, \ and\ \bibinfo {author} {\bibfnamefont {K.}~\bibnamefont
  {Horodecki}},\ }\bibfield  {title} {\enquote {\bibinfo {title} {Quantum
  entanglement},}\ }\href {https://doi.org/10.1103/RevModPhys.81.865}
  {\bibfield  {journal} {\bibinfo  {journal} {Rev. Mod. Phys.}\ }\textbf
  {\bibinfo {volume} {81}},\ \bibinfo {pages} {865} (\bibinfo {year}
  {2009})}\BibitemShut {NoStop}%
\bibitem [{\citenamefont {Bennett}\ \emph {et~al.}(1999)\citenamefont
  {Bennett}, \citenamefont {DiVincenzo}, \citenamefont {Fuchs}, \citenamefont
  {Mor}, \citenamefont {Rains}, \citenamefont {Shor}, \citenamefont {Smolin},\
  and\ \citenamefont {Wootters}}]{bennett1999quantum}%
  \BibitemOpen
  \bibfield  {author} {\bibinfo {author} {\bibfnamefont {Charles~H}\
  \bibnamefont {Bennett}}, \bibinfo {author} {\bibfnamefont {David~P}\
  \bibnamefont {DiVincenzo}}, \bibinfo {author} {\bibfnamefont {Christopher~A}\
  \bibnamefont {Fuchs}}, \bibinfo {author} {\bibfnamefont {Tal}\ \bibnamefont
  {Mor}}, \bibinfo {author} {\bibfnamefont {Eric}\ \bibnamefont {Rains}},
  \bibinfo {author} {\bibfnamefont {Peter~W}\ \bibnamefont {Shor}}, \bibinfo
  {author} {\bibfnamefont {John~A}\ \bibnamefont {Smolin}}, \ and\ \bibinfo
  {author} {\bibfnamefont {William~K}\ \bibnamefont {Wootters}},\ }\bibfield
  {title} {\enquote {\bibinfo {title} {Quantum nonlocality without
  entanglement},}\ }\href {https://doi.org/10.1103/PhysRevA.59.1070} {\bibfield
   {journal} {\bibinfo  {journal} {Physical Review A}\ }\textbf {\bibinfo
  {volume} {59}},\ \bibinfo {pages} {1070} (\bibinfo {year}
  {1999})}\BibitemShut {NoStop}%
\bibitem [{\citenamefont {Ollivier}\ and\ \citenamefont
  {Zurek}(2001)}]{ollivier2001quantum}%
  \BibitemOpen
  \bibfield  {author} {\bibinfo {author} {\bibfnamefont {H.}~\bibnamefont
  {Ollivier}}\ and\ \bibinfo {author} {\bibfnamefont {W.~H.}\ \bibnamefont
  {Zurek}},\ }\bibfield  {title} {\enquote {\bibinfo {title} {Quantum discord:
  a measure of the quantumness of correlations},}\ }\href
  {https://doi.org/10.1103/PhysRevLett.88.017901} {\bibfield  {journal}
  {\bibinfo  {journal} {Phys. Rev. Lett.}\ }\textbf {\bibinfo {volume} {88}},\
  \bibinfo {pages} {017901} (\bibinfo {year} {2001})}\BibitemShut {NoStop}%
\bibitem [{\citenamefont {Henderson}\ and\ \citenamefont
  {Vedral}(2001)}]{henderson2001classical}%
  \BibitemOpen
  \bibfield  {author} {\bibinfo {author} {\bibfnamefont {L.}~\bibnamefont
  {Henderson}}\ and\ \bibinfo {author} {\bibfnamefont {V.}~\bibnamefont
  {Vedral}},\ }\bibfield  {title} {\enquote {\bibinfo {title} {Classical,
  quantum and total correlations},}\ }\href
  {https://doi.org/10.1088/0305-4470/34/35/315} {\bibfield  {journal} {\bibinfo
   {journal} {J. Phys. A: Math. Gen.}\ }\textbf {\bibinfo {volume} {34}},\
  \bibinfo {pages} {6899} (\bibinfo {year} {2001})}\BibitemShut {NoStop}%
\bibitem [{\citenamefont {Maziero}\ \emph {et~al.}(2010)\citenamefont
  {Maziero}, \citenamefont {Celeri},\ and\ \citenamefont
  {Serra}}]{maziero2010symmetry}%
  \BibitemOpen
  \bibfield  {author} {\bibinfo {author} {\bibfnamefont {J.}~\bibnamefont
  {Maziero}}, \bibinfo {author} {\bibfnamefont {L.~C.}\ \bibnamefont {Celeri}},
  \ and\ \bibinfo {author} {\bibfnamefont {R.~M.}\ \bibnamefont {Serra}},\
  }\bibfield  {title} {\enquote {\bibinfo {title} {Symmetry aspects of quantum
  discord},}\ }\href {https://arxiv.org/abs/1004.2082} {\bibfield  {journal}
  {\bibinfo  {journal} {arXiv preprint arXiv:1004.2082}\ } (\bibinfo {year}
  {2010})}\BibitemShut {NoStop}%
\bibitem [{\citenamefont {Piani}\ \emph {et~al.}(2008)\citenamefont {Piani},
  \citenamefont {Horodecki},\ and\ \citenamefont {Horodecki}}]{piani2008no}%
  \BibitemOpen
  \bibfield  {author} {\bibinfo {author} {\bibfnamefont {M.}~\bibnamefont
  {Piani}}, \bibinfo {author} {\bibfnamefont {P.}~\bibnamefont {Horodecki}}, \
  and\ \bibinfo {author} {\bibfnamefont {R.}~\bibnamefont {Horodecki}},\
  }\bibfield  {title} {\enquote {\bibinfo {title} {No-local-broadcasting
  theorem for multipartite quantum correlations},}\ }\href
  {https://doi.org/10.1103/PhysRevLett.100.090502} {\bibfield  {journal}
  {\bibinfo  {journal} {Phys. Rev. Lett.}\ }\textbf {\bibinfo {volume} {100}},\
  \bibinfo {pages} {090502} (\bibinfo {year} {2008})}\BibitemShut {NoStop}%
\bibitem [{\citenamefont {Wu}\ \emph {et~al.}(2009)\citenamefont {Wu},
  \citenamefont {Poulsen},\ and\ \citenamefont
  {M\o{}lmer}}]{wu2009correlations}%
  \BibitemOpen
  \bibfield  {author} {\bibinfo {author} {\bibfnamefont {S.}~\bibnamefont
  {Wu}}, \bibinfo {author} {\bibfnamefont {U.~V.}\ \bibnamefont {Poulsen}}, \
  and\ \bibinfo {author} {\bibfnamefont {K.}~\bibnamefont {M\o{}lmer}},\
  }\bibfield  {title} {\enquote {\bibinfo {title} {Correlations in local
  measurements on a quantum state, and complementarity as an explanation of
  nonclassicality},}\ }\href {https://doi.org/10.1103/PhysRevA.80.032319}
  {\bibfield  {journal} {\bibinfo  {journal} {Phys. Rev. A}\ }\textbf {\bibinfo
  {volume} {80}},\ \bibinfo {pages} {032319} (\bibinfo {year}
  {2009})}\BibitemShut {NoStop}%
\bibitem [{\citenamefont {Terhal}\ \emph {et~al.}(2002)\citenamefont {Terhal},
  \citenamefont {Horodecki}, \citenamefont {Leung},\ and\ \citenamefont
  {DiVincenzo}}]{terhal2002entanglement}%
  \BibitemOpen
  \bibfield  {author} {\bibinfo {author} {\bibfnamefont {B.~M.}\ \bibnamefont
  {Terhal}}, \bibinfo {author} {\bibfnamefont {M.}~\bibnamefont {Horodecki}},
  \bibinfo {author} {\bibfnamefont {D.~W.}\ \bibnamefont {Leung}}, \ and\
  \bibinfo {author} {\bibfnamefont {D.~P.}\ \bibnamefont {DiVincenzo}},\
  }\bibfield  {title} {\enquote {\bibinfo {title} {The entanglement of
  purification},}\ }\href {https://doi.org/10.1063/1.1498001} {\bibfield
  {journal} {\bibinfo  {journal} {J. Math. Phys.}\ }\textbf {\bibinfo {volume}
  {43}},\ \bibinfo {pages} {4286--4298} (\bibinfo {year} {2002})}\BibitemShut
  {NoStop}%
\bibitem [{\citenamefont {Rulli}\ and\ \citenamefont
  {Sarandy}(2011)}]{rulli2011global}%
  \BibitemOpen
  \bibfield  {author} {\bibinfo {author} {\bibfnamefont {C.~C.}\ \bibnamefont
  {Rulli}}\ and\ \bibinfo {author} {\bibfnamefont {M.~S.}\ \bibnamefont
  {Sarandy}},\ }\bibfield  {title} {\enquote {\bibinfo {title} {Global quantum
  discord in multipartite systems},}\ }\href
  {https://doi.org/10.1103/PhysRevA.84.042109} {\bibfield  {journal} {\bibinfo
  {journal} {Phys. Rev. A}\ }\textbf {\bibinfo {volume} {84}},\ \bibinfo
  {pages} {042109} (\bibinfo {year} {2011})}\BibitemShut {NoStop}%
\bibitem [{\citenamefont {Huang}(2014)}]{1367-2630-16-3-033027}%
  \BibitemOpen
  \bibfield  {author} {\bibinfo {author} {\bibfnamefont {Yichen}\ \bibnamefont
  {Huang}},\ }\bibfield  {title} {\enquote {\bibinfo {title} {Computing quantum
  discord is {NP}-complete},}\ }\href
  {http://stacks.iop.org/1367-2630/16/i=3/a=033027} {\bibfield  {journal}
  {\bibinfo  {journal} {New J. Phys.}\ }\textbf {\bibinfo {volume} {16}},\
  \bibinfo {pages} {033027} (\bibinfo {year} {2014})}\BibitemShut {NoStop}%
\bibitem [{\citenamefont {Giorda}\ and\ \citenamefont
  {Paris}(2010)}]{giorda2010gaussian}%
  \BibitemOpen
  \bibfield  {author} {\bibinfo {author} {\bibfnamefont {P.}~\bibnamefont
  {Giorda}}\ and\ \bibinfo {author} {\bibfnamefont {M.~G.~A.}\ \bibnamefont
  {Paris}},\ }\bibfield  {title} {\enquote {\bibinfo {title} {Gaussian quantum
  discord},}\ }\href {https://doi.org/10.1103/PhysRevLett.105.020503}
  {\bibfield  {journal} {\bibinfo  {journal} {Phys. Rev. Lett.}\ }\textbf
  {\bibinfo {volume} {105}},\ \bibinfo {pages} {020503} (\bibinfo {year}
  {2010})}\BibitemShut {NoStop}%
\bibitem [{\citenamefont {Adesso}\ and\ \citenamefont
  {Datta}(2010)}]{adesso2010quantum}%
  \BibitemOpen
  \bibfield  {author} {\bibinfo {author} {\bibfnamefont {G.}~\bibnamefont
  {Adesso}}\ and\ \bibinfo {author} {\bibfnamefont {A.}~\bibnamefont {Datta}},\
  }\bibfield  {title} {\enquote {\bibinfo {title} {Quantum versus classical
  correlations in gaussian states},}\ }\href
  {https://link.aps.org/doi/10.1103/PhysRevLett.105.030501} {\bibfield
  {journal} {\bibinfo  {journal} {Phys. Rev. Lett.}\ }\textbf {\bibinfo
  {volume} {105}},\ \bibinfo {pages} {030501} (\bibinfo {year}
  {2010})}\BibitemShut {NoStop}%
\bibitem [{\citenamefont {Modi}\ \emph {et~al.}(2010)\citenamefont {Modi},
  \citenamefont {Paterek}, \citenamefont {Son}, \citenamefont {Vedral},\ and\
  \citenamefont {Williamson}}]{modi2010unified}%
  \BibitemOpen
  \bibfield  {author} {\bibinfo {author} {\bibfnamefont {K.}~\bibnamefont
  {Modi}}, \bibinfo {author} {\bibfnamefont {T.}~\bibnamefont {Paterek}},
  \bibinfo {author} {\bibfnamefont {W.}~\bibnamefont {Son}}, \bibinfo {author}
  {\bibfnamefont {V.}~\bibnamefont {Vedral}}, \ and\ \bibinfo {author}
  {\bibfnamefont {M.}~\bibnamefont {Williamson}},\ }\bibfield  {title}
  {\enquote {\bibinfo {title} {Unified view of quantum and classical
  correlations},}\ }\href {https://doi.org/10.1103/PhysRevLett.104.080501}
  {\bibfield  {journal} {\bibinfo  {journal} {Phys. Rev. Lett.}\ }\textbf
  {\bibinfo {volume} {104}},\ \bibinfo {pages} {080501} (\bibinfo {year}
  {2010})}\BibitemShut {NoStop}%
\bibitem [{\citenamefont {Daki\ifmmode~\acute{c}\else \'{c}\fi{}}\ \emph
  {et~al.}(2010)\citenamefont {Daki\ifmmode~\acute{c}\else \'{c}\fi{}},
  \citenamefont {Vedral},\ and\ \citenamefont {Brukner}}]{dakic2010necessary}%
  \BibitemOpen
  \bibfield  {author} {\bibinfo {author} {\bibfnamefont {B.}~\bibnamefont
  {Daki\ifmmode~\acute{c}\else \'{c}\fi{}}}, \bibinfo {author} {\bibfnamefont
  {V.}~\bibnamefont {Vedral}}, \ and\ \bibinfo {author} {\bibfnamefont
  {{\v{C}}.}~\bibnamefont {Brukner}},\ }\bibfield  {title} {\enquote {\bibinfo
  {title} {Necessary and sufficient condition for nonzero quantum discord},}\
  }\href {https://doi.org/10.1103/PhysRevLett.105.190502} {\bibfield  {journal}
  {\bibinfo  {journal} {Phys. Rev. Lett.}\ }\textbf {\bibinfo {volume} {105}},\
  \bibinfo {pages} {190502} (\bibinfo {year} {2010})}\BibitemShut {NoStop}%
\bibitem [{\citenamefont {Oppenheim}\ \emph {et~al.}(2002)\citenamefont
  {Oppenheim}, \citenamefont {Horodecki}, \citenamefont {Horodecki},\ and\
  \citenamefont {Horodecki}}]{oppenheim2002thermodynamical}%
  \BibitemOpen
  \bibfield  {author} {\bibinfo {author} {\bibfnamefont {J.}~\bibnamefont
  {Oppenheim}}, \bibinfo {author} {\bibfnamefont {M.}~\bibnamefont
  {Horodecki}}, \bibinfo {author} {\bibfnamefont {P.}~\bibnamefont
  {Horodecki}}, \ and\ \bibinfo {author} {\bibfnamefont {R.}~\bibnamefont
  {Horodecki}},\ }\bibfield  {title} {\enquote {\bibinfo {title}
  {Thermodynamical approach to quantifying quantum correlations},}\ }\href
  {https://doi.org/10.1103/PhysRevLett.89.180402} {\bibfield  {journal}
  {\bibinfo  {journal} {Phys. Rev. Lett.}\ }\textbf {\bibinfo {volume} {89}},\
  \bibinfo {pages} {180402} (\bibinfo {year} {2002})}\BibitemShut {NoStop}%
\bibitem [{\citenamefont {Luo}\ and\ \citenamefont
  {Fu}(2011)}]{luo2011measurement}%
  \BibitemOpen
  \bibfield  {author} {\bibinfo {author} {\bibfnamefont {S.}~\bibnamefont
  {Luo}}\ and\ \bibinfo {author} {\bibfnamefont {S.}~\bibnamefont {Fu}},\
  }\bibfield  {title} {\enquote {\bibinfo {title} {Measurement-induced
  nonlocality},}\ }\href {https://doi.org/10.1103/PhysRevLett.106.120401}
  {\bibfield  {journal} {\bibinfo  {journal} {Phys. Rev. Lett.}\ }\textbf
  {\bibinfo {volume} {106}},\ \bibinfo {pages} {120401} (\bibinfo {year}
  {2011})}\BibitemShut {NoStop}%
\bibitem [{\citenamefont {Girolami}\ \emph {et~al.}(2014)\citenamefont
  {Girolami}, \citenamefont {Souza}, \citenamefont {Giovannetti}, \citenamefont
  {Tufarelli}, \citenamefont {Filgueiras}, \citenamefont {Sarthour},
  \citenamefont {Soares-Pinto}, \citenamefont {Oliveira},\ and\ \citenamefont
  {Adesso}}]{qantum2014girolami}%
  \BibitemOpen
  \bibfield  {author} {\bibinfo {author} {\bibfnamefont {D.}~\bibnamefont
  {Girolami}}, \bibinfo {author} {\bibfnamefont {A.~M.}\ \bibnamefont {Souza}},
  \bibinfo {author} {\bibfnamefont {V.}~\bibnamefont {Giovannetti}}, \bibinfo
  {author} {\bibfnamefont {T.}~\bibnamefont {Tufarelli}}, \bibinfo {author}
  {\bibfnamefont {J.~G.}\ \bibnamefont {Filgueiras}}, \bibinfo {author}
  {\bibfnamefont {R.~S.}\ \bibnamefont {Sarthour}}, \bibinfo {author}
  {\bibfnamefont {D.~O.}\ \bibnamefont {Soares-Pinto}}, \bibinfo {author}
  {\bibfnamefont {I.~S.}\ \bibnamefont {Oliveira}}, \ and\ \bibinfo {author}
  {\bibfnamefont {G.}~\bibnamefont {Adesso}},\ }\bibfield  {title} {\enquote
  {\bibinfo {title} {Quantum discord determines the interferometric power of
  quantum states},}\ }\href
  {https://link.aps.org/doi/10.1103/PhysRevLett.112.210401} {\bibfield
  {journal} {\bibinfo  {journal} {Phys. Rev. Lett.}\ }\textbf {\bibinfo
  {volume} {112}},\ \bibinfo {pages} {210401} (\bibinfo {year}
  {2014})}\BibitemShut {NoStop}%
\bibitem [{\citenamefont {Bera}\ \emph {et~al.}(2018)\citenamefont {Bera},
  \citenamefont {Das}, \citenamefont {Sadhukhan}, \citenamefont {Roy},
  \citenamefont {Sen(De)},\ and\ \citenamefont {Sen}}]{quantum2017bera}%
  \BibitemOpen
  \bibfield  {author} {\bibinfo {author} {\bibfnamefont {A.}~\bibnamefont
  {Bera}}, \bibinfo {author} {\bibfnamefont {T.}~\bibnamefont {Das}}, \bibinfo
  {author} {\bibfnamefont {D.}~\bibnamefont {Sadhukhan}}, \bibinfo {author}
  {\bibfnamefont {S.~S.}\ \bibnamefont {Roy}}, \bibinfo {author} {\bibfnamefont
  {A.}~\bibnamefont {Sen(De)}}, \ and\ \bibinfo {author} {\bibfnamefont
  {U.}~\bibnamefont {Sen}},\ }\bibfield  {title} {\enquote {\bibinfo {title}
  {Quantum discord and its allies: a review of recent progress},}\ }\href
  {https://doi.org/10.1088/1361-6633/aa872f} {\bibfield  {journal} {\bibinfo
  {journal} {Reports on Progress in Physics}\ }\textbf {\bibinfo {volume}
  {81}},\ \bibinfo {pages} {024001} (\bibinfo {year} {2018})}\BibitemShut
  {NoStop}%
\bibitem [{\citenamefont {Devetak}\ and\ \citenamefont
  {Winter}(2003)}]{devetak2003classical}%
  \BibitemOpen
  \bibfield  {author} {\bibinfo {author} {\bibfnamefont {I.}~\bibnamefont
  {Devetak}}\ and\ \bibinfo {author} {\bibfnamefont {A.}~\bibnamefont
  {Winter}},\ }\bibfield  {title} {\enquote {\bibinfo {title} {Classical data
  compression with quantum side information},}\ }\href
  {https://doi.org/10.1103/PhysRevA.68.042301} {\bibfield  {journal} {\bibinfo
  {journal} {Phys. Rev. A}\ }\textbf {\bibinfo {volume} {68}},\ \bibinfo
  {pages} {042301} (\bibinfo {year} {2003})}\BibitemShut {NoStop}%
\bibitem [{\citenamefont {Watanabe}(1960)}]{watanabe1960information}%
  \BibitemOpen
  \bibfield  {author} {\bibinfo {author} {\bibfnamefont {S.}~\bibnamefont
  {Watanabe}},\ }\bibfield  {title} {\enquote {\bibinfo {title} {Information
  theoretical analysis of multivariate correlation},}\ }\href
  {https://doi.org/10.1147/rd.41.0066} {\bibfield  {journal} {\bibinfo
  {journal} {IBM Journal of research and development}\ }\textbf {\bibinfo
  {volume} {4}},\ \bibinfo {pages} {66--82} (\bibinfo {year}
  {1960})}\BibitemShut {NoStop}%
\bibitem [{\citenamefont {Weedbrook}\ \emph {et~al.}(2012)\citenamefont
  {Weedbrook}, \citenamefont {Pirandola}, \citenamefont {Garc\'{\i}a-Patr\'on},
  \citenamefont {Cerf}, \citenamefont {Ralph}, \citenamefont {Shapiro},\ and\
  \citenamefont {Lloyd}}]{weedbrook2012gaussian}%
  \BibitemOpen
  \bibfield  {author} {\bibinfo {author} {\bibfnamefont {C.}~\bibnamefont
  {Weedbrook}}, \bibinfo {author} {\bibfnamefont {S.}~\bibnamefont
  {Pirandola}}, \bibinfo {author} {\bibfnamefont {R.}~\bibnamefont
  {Garc\'{\i}a-Patr\'on}}, \bibinfo {author} {\bibfnamefont {N.~J.}\
  \bibnamefont {Cerf}}, \bibinfo {author} {\bibfnamefont {T.~C.}\ \bibnamefont
  {Ralph}}, \bibinfo {author} {\bibfnamefont {J.~H.}\ \bibnamefont {Shapiro}},
  \ and\ \bibinfo {author} {\bibfnamefont {S.}~\bibnamefont {Lloyd}},\
  }\bibfield  {title} {\enquote {\bibinfo {title} {Gaussian quantum
  information},}\ }\href {http://dx.doi.org/10.1103/RevModPhys.84.621}
  {\bibfield  {journal} {\bibinfo  {journal} {Rev. Mod. Phys.}\ }\textbf
  {\bibinfo {volume} {84}},\ \bibinfo {pages} {621--669} (\bibinfo {year}
  {2012})}\BibitemShut {NoStop}%
\bibitem [{\citenamefont {Mi{\v{s}}ta~Jr.}\ and\ \citenamefont
  {Tatham}(2016)}]{mivsta2016gaussian}%
  \BibitemOpen
  \bibfield  {author} {\bibinfo {author} {\bibfnamefont {L.}~\bibnamefont
  {Mi{\v{s}}ta~Jr.}}\ and\ \bibinfo {author} {\bibfnamefont {R.}~\bibnamefont
  {Tatham}},\ }\bibfield  {title} {\enquote {\bibinfo {title} {Gaussian
  intrinsic entanglement},}\ }\href
  {https://doi.org/10.1103/PhysRevLett.117.240505} {\bibfield  {journal}
  {\bibinfo  {journal} {Phys. Rev. Lett.}\ }\textbf {\bibinfo {volume} {117}},\
  \bibinfo {pages} {240505} (\bibinfo {year} {2016})}\BibitemShut {NoStop}%
\bibitem [{\citenamefont {Duan}\ \emph {et~al.}(2000)\citenamefont {Duan},
  \citenamefont {Giedke}, \citenamefont {Cirac},\ and\ \citenamefont
  {Zoller}}]{duan2000inseparability}%
  \BibitemOpen
  \bibfield  {author} {\bibinfo {author} {\bibfnamefont {L.-M.}\ \bibnamefont
  {Duan}}, \bibinfo {author} {\bibfnamefont {G.}~\bibnamefont {Giedke}},
  \bibinfo {author} {\bibfnamefont {J.~I.}\ \bibnamefont {Cirac}}, \ and\
  \bibinfo {author} {\bibfnamefont {P.}~\bibnamefont {Zoller}},\ }\bibfield
  {title} {\enquote {\bibinfo {title} {Inseparability criterion for continuous
  variable systems},}\ }\href {https://doi.org/10.1103/PhysRevLett.84.2722}
  {\bibfield  {journal} {\bibinfo  {journal} {Phys. Rev. Lett.}\ }\textbf
  {\bibinfo {volume} {84}},\ \bibinfo {pages} {2722} (\bibinfo {year}
  {2000})}\BibitemShut {NoStop}%
\bibitem [{\citenamefont {Pirandola}\ \emph {et~al.}(2014)\citenamefont
  {Pirandola}, \citenamefont {Spedalieri}, \citenamefont {Braunstein},
  \citenamefont {Cerf},\ and\ \citenamefont {Lloyd}}]{pirandola2014optimality}%
  \BibitemOpen
  \bibfield  {author} {\bibinfo {author} {\bibfnamefont {S.}~\bibnamefont
  {Pirandola}}, \bibinfo {author} {\bibfnamefont {G.}~\bibnamefont
  {Spedalieri}}, \bibinfo {author} {\bibfnamefont {S.~L.}\ \bibnamefont
  {Braunstein}}, \bibinfo {author} {\bibfnamefont {N.~J.}\ \bibnamefont
  {Cerf}}, \ and\ \bibinfo {author} {\bibfnamefont {S.}~\bibnamefont {Lloyd}},\
  }\bibfield  {title} {\enquote {\bibinfo {title} {Optimality of gaussian
  discord},}\ }\href {https://doi.org/10.1103/PhysRevLett.113.140405}
  {\bibfield  {journal} {\bibinfo  {journal} {Phys. Rev. Lett.}\ }\textbf
  {\bibinfo {volume} {113}},\ \bibinfo {pages} {140405} (\bibinfo {year}
  {2014})}\BibitemShut {NoStop}%
\bibitem [{\citenamefont {Gu}\ \emph {et~al.}(2012)\citenamefont {Gu},
  \citenamefont {Chrzanowski}, \citenamefont {Assad}, \citenamefont {Symul},
  \citenamefont {Modi}, \citenamefont {Ralph}, \citenamefont {Vedral},\ and\
  \citenamefont {Lam}}]{gu2012observing}%
  \BibitemOpen
  \bibfield  {author} {\bibinfo {author} {\bibfnamefont {Mile}\ \bibnamefont
  {Gu}}, \bibinfo {author} {\bibfnamefont {Helen~M}\ \bibnamefont
  {Chrzanowski}}, \bibinfo {author} {\bibfnamefont {Syed~M}\ \bibnamefont
  {Assad}}, \bibinfo {author} {\bibfnamefont {Thomas}\ \bibnamefont {Symul}},
  \bibinfo {author} {\bibfnamefont {Kavan}\ \bibnamefont {Modi}}, \bibinfo
  {author} {\bibfnamefont {Timothy~C}\ \bibnamefont {Ralph}}, \bibinfo {author}
  {\bibfnamefont {Vlatko}\ \bibnamefont {Vedral}}, \ and\ \bibinfo {author}
  {\bibfnamefont {Ping~Koy}\ \bibnamefont {Lam}},\ }\bibfield  {title}
  {\enquote {\bibinfo {title} {Observing the operational significance of
  discord consumption},}\ }\href {https://doi.org/10.1038/nphys2376} {\bibfield
   {journal} {\bibinfo  {journal} {Nature Physics}\ }\textbf {\bibinfo {volume}
  {8}},\ \bibinfo {pages} {671} (\bibinfo {year} {2012})}\BibitemShut {NoStop}%
\bibitem [{\citenamefont {Adesso}\ \emph {et~al.}(2006)\citenamefont {Adesso},
  \citenamefont {Serafini},\ and\ \citenamefont
  {Illuminati}}]{adesso2006multipartite}%
  \BibitemOpen
  \bibfield  {author} {\bibinfo {author} {\bibfnamefont {G.}~\bibnamefont
  {Adesso}}, \bibinfo {author} {\bibfnamefont {A.}~\bibnamefont {Serafini}}, \
  and\ \bibinfo {author} {\bibfnamefont {F.}~\bibnamefont {Illuminati}},\
  }\bibfield  {title} {\enquote {\bibinfo {title} {Multipartite entanglement in
  three-mode gaussian states of continuous-variable systems: Quantification,
  sharing structure, and decoherence},}\ }\href
  {https://doi.org/10.1103/PhysRevA.73.032345} {\bibfield  {journal} {\bibinfo
  {journal} {Phys. Rev. A}\ }\textbf {\bibinfo {volume} {73}},\ \bibinfo
  {pages} {032345} (\bibinfo {year} {2006})}\BibitemShut {NoStop}%
\bibitem [{\citenamefont {Giedke}\ \emph {et~al.}(2001)\citenamefont {Giedke},
  \citenamefont {Kraus}, \citenamefont {Lewenstein},\ and\ \citenamefont
  {Cirac}}]{giedke2001separability}%
  \BibitemOpen
  \bibfield  {author} {\bibinfo {author} {\bibfnamefont {G.}~\bibnamefont
  {Giedke}}, \bibinfo {author} {\bibfnamefont {B.}~\bibnamefont {Kraus}},
  \bibinfo {author} {\bibfnamefont {M.}~\bibnamefont {Lewenstein}}, \ and\
  \bibinfo {author} {\bibfnamefont {J.~I.}\ \bibnamefont {Cirac}},\ }\bibfield
  {title} {\enquote {\bibinfo {title} {Separability properties of three-mode
  gaussian states},}\ }\href {\doibase 10.1103/PhysRevA.64.052303} {\bibfield
  {journal} {\bibinfo  {journal} {Phys. Rev. A}\ }\textbf {\bibinfo {volume}
  {64}},\ \bibinfo {pages} {052303} (\bibinfo {year} {2001})}\BibitemShut
  {NoStop}%
\end{thebibliography}%

\end{document}